\documentclass[11pt]{article}

\usepackage{ epsfig,subfigure}
\usepackage{amssymb, amsmath}

\begin{document}
\newtheorem{theorem}{Theorem}
\newtheorem{corollary}{Corollary}
\newtheorem{definition}{Definition}
\newtheorem{lemma}{Lemma}

\newcommand{\define}{\stackrel{\triangle}{=}}
\def\QED{\mbox{\rule[0pt]{1.5ex}{1.5ex}}}
\def\proof{\noindent\hspace{2em}{\it Proof: }}

\date{}

\title{Optimal Use of Current and  Outdated  Channel State Information --- Degrees of Freedom of the MISO BC with Mixed CSIT}
\author{ Tiangao Gou and Syed A. Jafar\\
Center for Pervasive Communications and Computing (CPCC)\\
University of California Irvine, Irvine, CA 92697-2625\\
{\small \it  Email~: \{tgou, syed\}@uci.edu}}

\maketitle

\thispagestyle{empty}
\begin{abstract}
We consider a multiple-input-single-output (MISO) broadcast channel with mixed channel state information at the transmitter (CSIT) that consists of imperfect current CSIT and perfect outdated CSIT. Recent work by Kobayashi et al. presented a scheme which exploits both imperfect current CSIT and perfect outdated CSIT and achieves higher degrees of freedom (DoF) than possible with only imperfect current CSIT or only outdated CSIT individually. In this work, we further improve the achievable DoF in this setting by incorporating additional private messages, and provide a tight information theoretic DoF outer bound, thereby identifying the DoF optimal use of mixed CSIT. The new result is stronger even in the original setting of only delayed CSIT, because it allows us to remove the restricting assumption of statistically equivalent fading for all users.
\end{abstract}

\section{Introduction}
Channel state information at transmitter (CSIT) is an important issue when designing communication systems, and can be available in a variety of forms. Consider the following CSIT models for a two user MISO broadcast channel (BC) where the transmitter is equipped with two antennas, each receiver has one antenna and the channels vary in an i.i.d. fashion across time.
\begin{enumerate}
\item {\bf Perfect current CSIT:} This is the setting where the transmitter knows the instantaneous channels perfectly at time $t$.
\item {\bf Delayed CSIT:} This is the setting where at time $t$, the transmitter knows the channels up to time $t-1$ perfectly.
\item {\bf Delayed CSIT and imperfect current CSIT:} This is the setting where the transmitter has delayed CSIT and also has partial knowledge of the channels at time $t$.
\end{enumerate}
When the channel is changing very slowly,  perfect current CSIT is a reasonable assumption. Under this model, zero forcing at the transmitter allows each user to achieve 1 DoF which is also its interference-free DoF. On the other hand, if the channel is changing very rapidly, delayed CSIT is a reasonable assumption. A channel that changes from symbol to symbol in an  i.i.d.fashion, making any CSIT completely outdated (i.e., independent of the current channel state), naturally represents a worst case scenario. Surprisingly, even in this case, a DoF gain can be obtained due to retrospective interference alignment \cite{Maddah_Tse}, \cite{Maleki_Jafar_Shamai}. For many practical settings, however,  at least some imperfect knowledge of current channel state may be available in addition to the past channel state information. This can happen, e.g.,  due to temporal channel correlations when the CSIT feedback is not too delayed, or, e.g., due to the availability  of  a  feedback channel  with significant delay (providing outdated CSIT)  in addition to the observations from a reverse channel (due to the two-way nature of communication) which may provide an (imperfect) estimate of the current channel state. In this work we  explore how the transmitter can \emph{optimally} use this mixed CSIT, from a DoF perspective.

\subsection{Overview of results}
We consider a two user MISO broadcast channel where the transmitter has two antennas. It is assumed that the transmitter has perfect delayed CSI and imperfect current CSI. We are interested in characterizing the DoF of this channel. To better understand the results, we present the channel model in its simplest form as follows. The rationale for this simple form  will be explained in detail in Section \ref{sec:simplemodel}.
\begin{figure}[!t]
\centering
\includegraphics[width=3.5in]{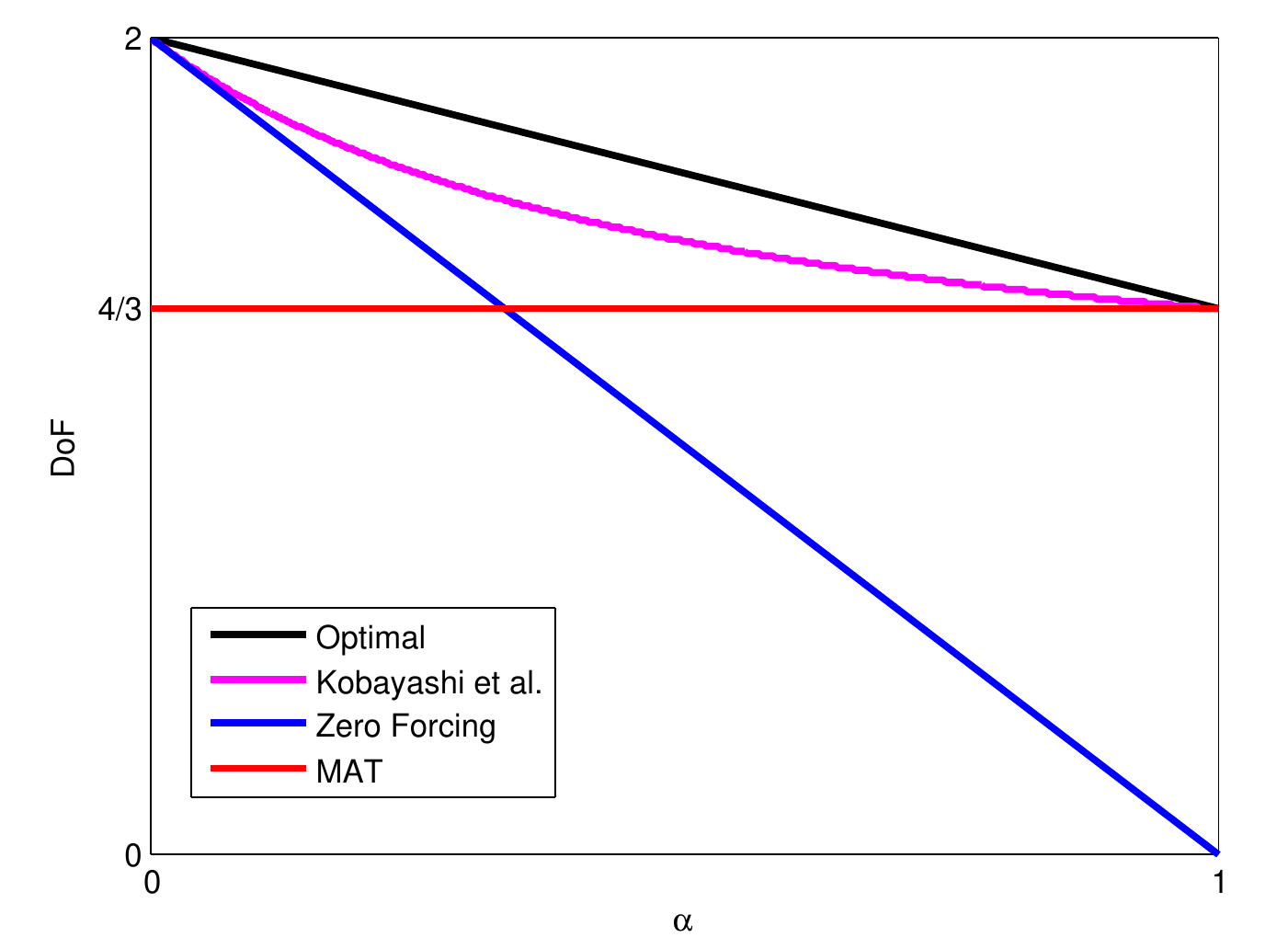}
\caption{DoF as a function of $\alpha$ }
\label{fig:dof}
\end{figure}
\begin{eqnarray}
y_1(t)&=&\sqrt{P}~x_1(t)+h(t)\sqrt{P^\alpha}~ x_2(t)+z_1(t)\label{eq:model1}\\
y_2(t)&=&\sqrt{P}~x_2(t)+g(t)\sqrt{P^\alpha} ~x_1(t)+z_2(t)\label{eq:model2}
\end{eqnarray}
Here, during channel use $t$, $y_i(t)$ is the received signal at receiver $i$, $z_i(t)$ is the additive white Gaussian noise with zero mean and finite variance at receiver $i$, $x_1(t)$ is the projection of the transmitted signal along the dimension that is orthogonal to the estimated current channel of receiver 2 while $x_2(t)$ is the projection of the transmitted signal along the dimension that is orthogonal to the estimated current channel of receiver 1. Essentially, by a change of basis operation (no loss of generality for DoF), the effective transmit antennas are aligned with the zero forcing signaling dimensions based on the imperfect current CSIT. Since the current channel is not known perfectly, perfect zero-forcing of signals is not possible, so that $h(t)$ and $g(t)$  are the channels for the non-zero-forced part of the signals. They are modeled as unknown channel coefficients at time $t$, and  will be known to the transmitter after a significant delay (outdated CSIT). Due to imperfect current CSIT, $P^{\alpha}$ represents the residual signal power after zero forcing, whose strength is measured by  $\alpha$, a parameter that ranges between zero and one. When $\alpha=0$, it corresponds to the case when current CSIT is as good as perfect, because the residual  signal power after zero forcing is at the noise floor  level and has no impact on DoF. In this case, simply zero-forcing allows each user to achieve 1 DoF, for a total of 2 DoF which is also the maximum DoF of the channel  even with perfect current CSIT. On the other hand, if $\alpha=1$, it corresponds to the case when there is no current CSIT, and zero-forcing is not able to reduce the signal strength. In this case,  Maddah-Ali and Tse proposed an interference alignment scheme (which will be referred to as the MAT scheme) to achieve the optimal DoF of $\frac{4}{3}$ \cite{Maddah_Tse}.

For $0<\alpha<1$, zero-forcing, which  only uses current CSI, achieves $2-2\alpha$ DoF, while the MAT scheme, which only uses delayed CSI, achieves $4/3$ DoF regardless of $\alpha$. Recently, in \cite{Kobayashi_Yang}, Kobayashi et al. proposed an interesting scheme which exploits both current and delayed CSI to achieve $\frac{2(1+\alpha)}{1+2\alpha}$ DoF\footnote{The parameter $\alpha$ used in this paper is different from that used in \cite{Kobayashi_Yang}. As will be explained in the following part of the paper, suppose we denote the $\alpha$ used in \cite{Kobayashi_Yang} as $\alpha'$. Then $\alpha$ in this paper is related to $\alpha'$ through $\alpha=1-\alpha'$. } \cite{Kobayashi_Yang}. The achievable DoF curve of \cite{Kobayashi_Yang} is shown in Fig. \ref{fig:dof}. In this work, we build our scheme upon that of Kobayashi et al. and further improve it to achieve $2-\frac{2}{3}\alpha$ DoF, which represents a straight line that goes from a sum DoF value of 2 when $\alpha=0$ to a sum DoF value of $4/3$ when $\alpha=1$ as shown in Fig. \ref{fig:dof}. The new achievable scheme provides the largest DoF gain of $4/9$ over best of ZF and MAT schemes at $\alpha=1/3$.

We also derive a DoF outer bound for the mixed CSIT setting, where no non-trivial outer bound was previously available, to show that $2-\frac{2}{3}\alpha$ DoF is in fact \emph{optimal}. In addition to being tight, the outer bound possesses another robust feature --- it  does not force the i.i.d. assumption across users, i.e., the users can have different channel distributions. In contrast, the assumption of statistically equivalent fading for all users is used for the original outer bound derived for the MISO broadcast channel with only delayed CSIT \cite{Maddah_Tse}. Therefore, apparently even in the original setting of only delayed CSIT, our result further strengthens the result of \cite{Maddah_Tse}.

\subsection{The achievable schemes: key ideas}
In this section, we provide an intuitive description of  our achievable scheme. Since our scheme is built upon that of  \cite{Kobayashi_Yang}  (which is built upon \cite{Maddah_Tse}), we first review the scheme in \cite{Kobayashi_Yang}.

\subsubsection{Achievable scheme of Kobayashi et al. \cite{Kobayashi_Yang}}
Following the principle of MAT scheme, in the scheme of Kobayashi et al. also, the interference initially seen at each receiver is multicast to both receivers in subsequent transmission phases, such that the undesired receiver can cancel the previously seen interference while the desired receiver can use this additional observation to resolve its desired symbols. However, unlike MAT scheme which assumes no knowledge of current channel state, Kobayashi et al. also simultaneously take advantage of the partial zero forcing capability provided by the imperfect current CSIT. Specifically, the scheme consists of three phases.

In the first phase, which consists of one time slot, the transmitter simultaneously sends four symbols $a_1, a_2, b_1, b_2$, where $a_1, a_2$ are intended for receiver 1 and $b_1,b_2$ are intended for receiver 2.
\begin{figure}[!h]
\centering
\includegraphics[width=3.5in]{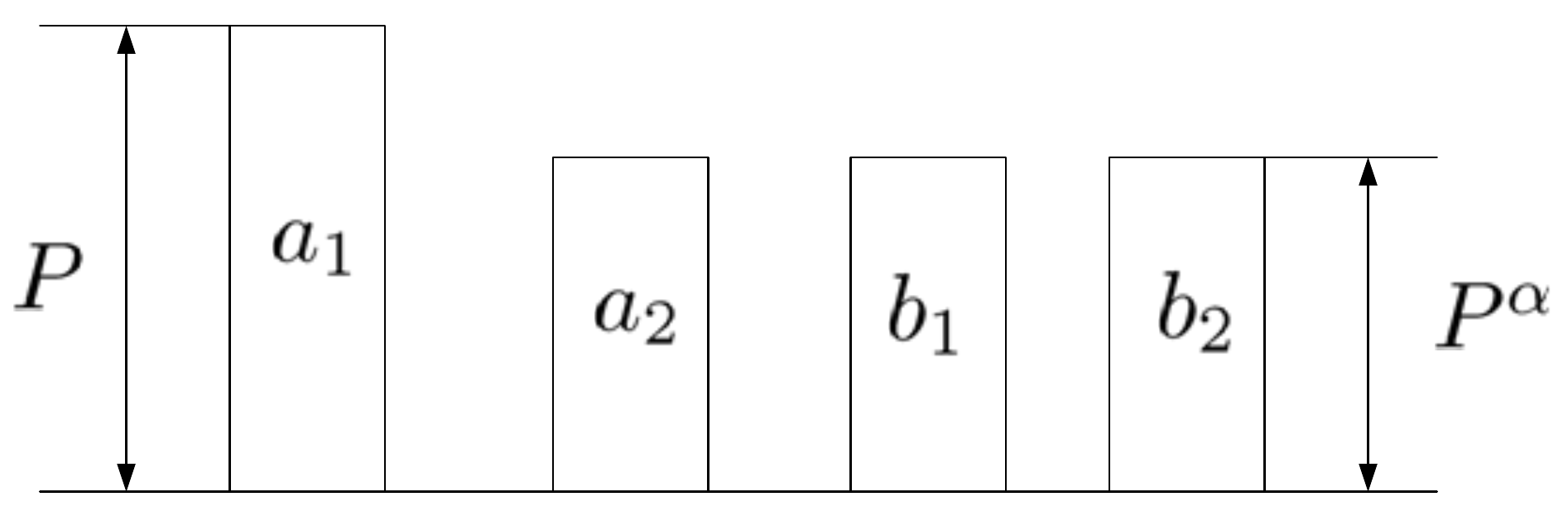}
\caption{Power levels of symbols at receiver 1 in the first time slot }
\label{fig:rx1t1_intro}
\end{figure}
The symbols $a_1$ and $b_1$ exploit the partial zero forcing capability, as they are sent along the directions that are orthogonal to the imperfectly known current channel states of receiver 2 and 1, respectively. Due to partial zero-forcing, $a_1, b_1$ are received with full power $P$ at the desired receivers, but only at power $P^{\alpha}$ at the undesired receivers. Having exhausted the partial zero-forcing capability with $a_1,b_1$, the transmitter simultaneously sends  $a_2$ and $b_2$ along generic directions to guarantee linear independence with $a_1, b_1$. However, $a_2, b_2$ are allocated  less power\footnote{According to the model (\ref{eq:model1}), (\ref{eq:model2}), the power allocated to $a_2, b_2$ is simply $P^{\alpha-1}$ so that the dominant received power term from these symbols is $P^\alpha$ at both receivers,  and the choice of generic directions  simply implies transmission from both antennas with different linear weights.} so that they are received at power level $P^\alpha$ at both receivers. Note that this is the same power level as $a_1$ and $b_1$ at the \emph{undesired} receivers. As shown in Fig. \ref{fig:rx1t1_intro}, $b_1$ and $b_2$ are received at the same power level $P^{\alpha}$ at receiver $1$, as is $a_2$.

In the next phase, the transmitter wants to deliver the linear combination of $b_1$ and $b_2$ seen by receiver 1 and that of $a_1$ and $a_2$ seen by receiver 2 to both receivers. A key novel idea of \cite{Kobayashi_Yang} is to \emph{quantize} the interference and then multicast it to both receivers digitally instead of directly sending the analog interference signal as the MAT scheme. Now consider quantizing the interference given by the linear combination of $b_1$ and $b_2$ at receiver 1. As shown in Fig. \ref{fig:rx1t1_intro}, since $b_1, b_2$ have power $P^{\alpha}$ at receiver 1, the linear combination can be quantized to within unit approximation error using approximately $\alpha \log P$ bits. These bits will be coded using a multicast code and then be sent to both receivers. Since the channel has 1 DoF for multicast to both receivers, i.e., rate $\log P$ for sending common information, the number of time slots needed to send these bits is approximately $\alpha$. 

Similarly, in the third phase the transmitter will use $\alpha$ time slots to multicast the quantized linear combination of $a_1$ and $a_2$ seen by receiver 2 in the first time slot.  As a consequence, combining all three phases, after $1+2\alpha$ time slots each user can decode two desired symbols to achieve $\frac{(1+\alpha)}{1+2\alpha}$ DoF.

\subsection{New (optimal) Scheme: Improving upon the achievable scheme of \cite{Kobayashi_Yang}}
\begin{figure*}[!h]
\centerline{\subfigure[Power levels of symbols at receiver 1
]{\includegraphics[width=2.2in]{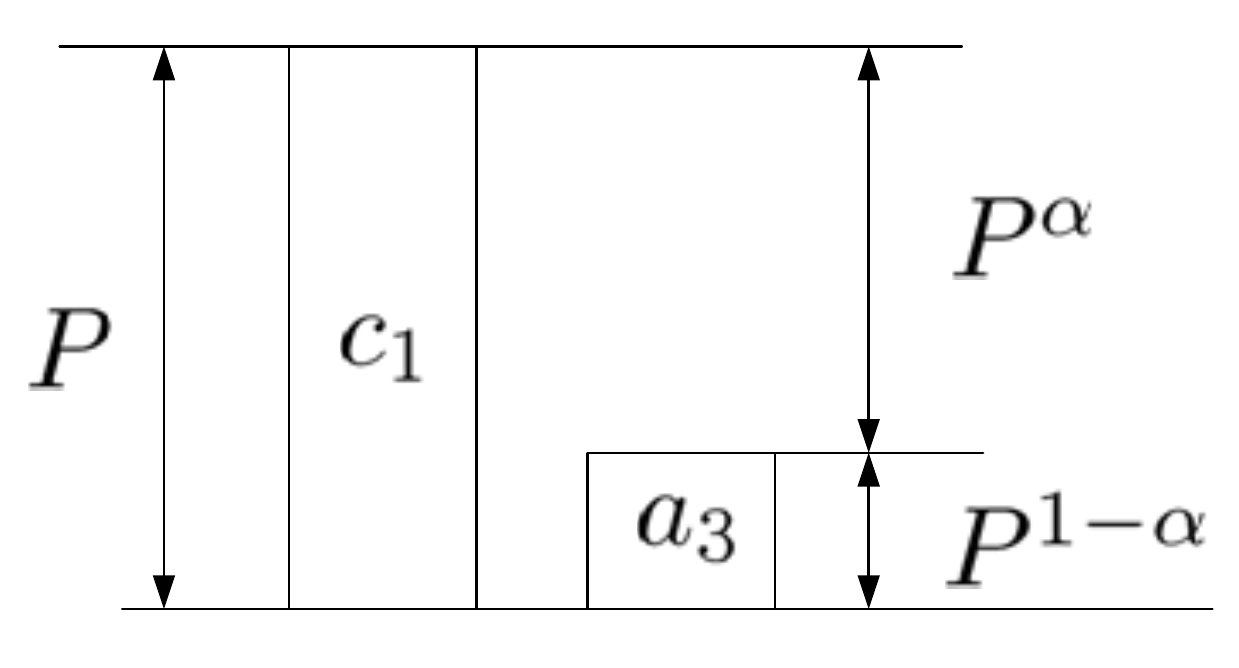} \label{fig:rx1t2_intro}} \hfil \subfigure[Power levels of symbols at receiver 2 ]{\includegraphics[width=2.2in]{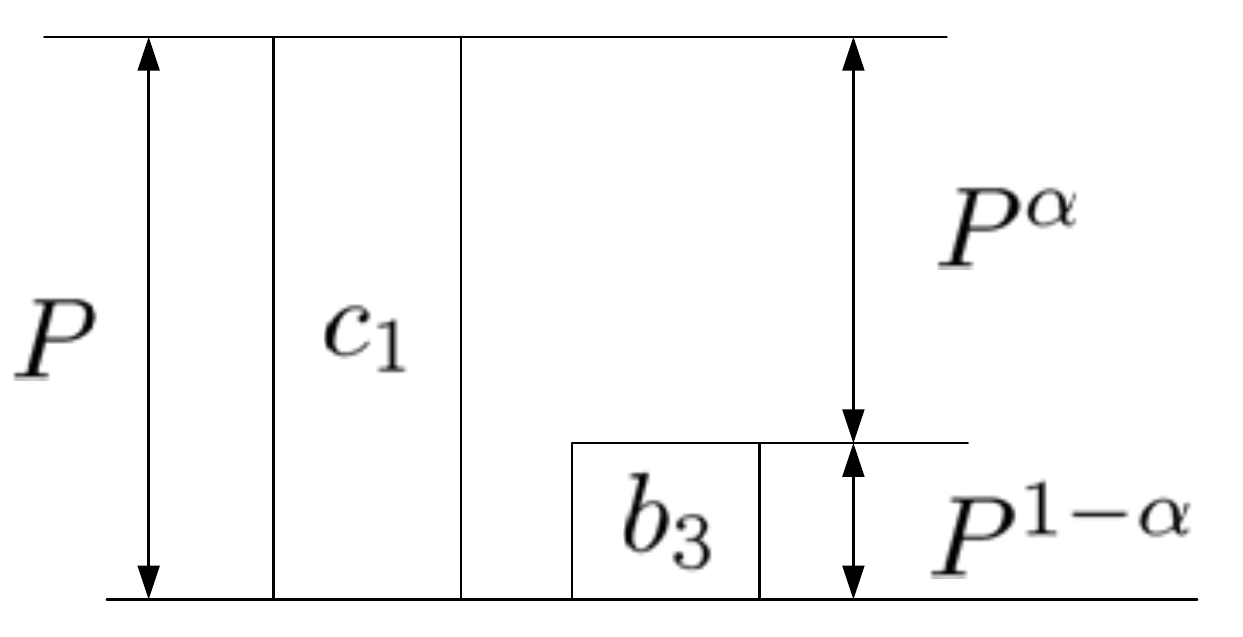}\label{fig:rx2t2_intro} }}
\caption{Our scheme: The power levels of symbols at receiver 1 and 2 in the second time slot}\label{fig:t2_intro}
\end{figure*}

Our scheme also consists of three phases. The main reason for the improvement is that while the scheme of Kobayashi et al.  uses the partial zero forcing capability of the channel only in Phase 1, our scheme uses the partial zero forcing capability in \emph{every} transmission. Phase 1 of our scheme is the same as that of Kobayashi et al. in \cite{Kobayashi_Yang}.  For the second phase, just like Kobayashi et al., we  also multicast the  common information (denoted by $c_1$ in Fig. \ref{fig:t2_intro}) which is simply the quantization of the linear combination of $b_1$ and $b_2$ seen by receiver 1. However, unlike the scheme of Kobayashi et al., in addition to multicasting this common information, \emph{we also simultaneously send two new private messages}, one for each receiver using the partial zero forcing capability of the channel. Denote these additional symbols for receiver 1 and 2 as $a_3$ and $b_3$, respectively. Then $a_3$ and $b_3$ are sent from $x_1$ and $x_2$, respectively, in \eqref{eq:model1} and $\eqref{eq:model2}$, i.e., they are sent along the imperfectly known zero-forcing directions. Moreover, we allocate only enough power to $a_3, b_3$  such that they are received at noise floor level at the unintended receivers. From \eqref{eq:model1} and $\eqref{eq:model2}$, it can be seen the allocated power is $P^{-\alpha}$. With this power allocation, they will be received at the power level $P^{1-\alpha}$ at the desired receivers. The power levels of symbols at two receivers are shown in Fig. \ref{fig:t2_intro}.
Each receiver will decode the common information $c_1$ first, subtract it out and then decode the private message. Due to the private message, the noise level for the common message is raised to $P^{1-\alpha}$. Thus, the common message can achieve a rate approximately $\log(P/P^{1-\alpha})=\alpha \log P$. Each private message can achieve a rate approximately $(1-\alpha)\log P$. Since the common message contains $\alpha \log P$ bits, only one time slot is need to deliver it. 

The third phase is similar to second phase, i.e., we send the common information, $c_2$, carrying the quantized linear combination of $a_1$ and $a_2$ as seen by receiver 2 in the first time slot, and two private messages, $a_4, b_4$,  one for each receiver. The remaining details are the same as phase 2.

As a consequence, over 3 time slots, each user can decode four desired symbols to achieve $\frac{1+\alpha+1-\alpha+1-\alpha}{3}=\frac{3-\alpha}{3}$ DoF. 

We also provide an information theoretic outer bound that proves this DoF value is optimal. The DoF value can be written as $2(\frac{2}{3}\alpha+1-\alpha)$ and can be interpreted as ``\emph{each user can achieve $\frac{2}{3}\alpha$ DoF  due to retrospective interference alignment and $1-\alpha$ DoF due to zero-forcing.}"

\section{System Model}\label{sec:systemmodel}
We follow a similar model to Kobayashi et al. \cite{Kobayashi_Yang}, except that we assume i.i.d. temporal variations.\footnote{The assumption of i.i.d. temporal variations is significant only for the DoF outer bound. For achievability, our scheme works perfectly with the same assumptions as Kobayashi et al. as well.} The channel input-output relationship at time $t$ is given by
\begin{eqnarray}\label{eqn:model}
y_1(t)&=&\mathbf{h}^{\dagger}(t)\mathbf{x}(t)+z_1(t)\\
y_2(t)&=&\mathbf{g}^{\dagger}(t)\mathbf{x}(t)+z_2(t)
\end{eqnarray}
where $y_i(t)$ is the observed signal at receiver $i$, $\mathbf{x}(t)=[x_1(t)~x_2(t)]^T$ is the $2\times 1$ input signal satisfying the power constraint $E[\|\mathbf{x}(t)\|^2]\leq P$ and $z_i(t)\sim \mathcal{CN}(0,1)$ is the circularly symmetric complex additive white Gaussian noise (AWGN). $\mathbf{h}(t)$ and $\mathbf{g}(t)$ are two $2\times 1$ channel vectors to receiver 1 and 2, respectively.  In addition, we assume that
\begin{eqnarray}
\mathbf{h}(t)=\hat{\mathbf{h}}(t)+\tilde{\mathbf{h}}(t)\\
\mathbf{g}(t)=\hat{\mathbf{g}}(t)+\tilde{\mathbf{g}}(t)
\end{eqnarray}
where $\hat{\mathbf{h}}(t)$ and $\hat{\mathbf{g}}(t)$ are the estimated channels while $\mathbf{\tilde{h}}(t)$ and $\tilde{\mathbf{g}}(t)$ are the estimation errors. We assume that $\hat{\mathbf{h}}(t), \tilde{\mathbf{h}}(t), \hat{\mathbf{g}}(t),\tilde{\mathbf{g}}(t)$ are drawn from continuous distributions, independent of each other (although not necessarily identically distributed), and vary in an i.i.d. fashion in time.

The estimated channels and estimation errors are assumed to be zero mean with covariance matrices $(1-\sigma^2)\mathbf{I}$ and $\sigma^2\mathbf{I}$ ($\sigma^2\leq 1$), respectively. Receivers know the instantaneous channels perfectly. Regarding the channel state information at the transmitter, it is assumed that the transmitter has 1) perfect delayed CSI and 2) imperfect current CSI, i.e., at time $t$ the transmitter knows perfectly $\mathbf{h}(1),\cdots,\mathbf{h}(t-1)$ and $\mathbf{g}(1),\cdots,\mathbf{g}(t-1)$ as well as $\hat{\mathbf{h}}(t)$ and $\hat{\mathbf{g}}(t)$. We define
\begin{equation}
\alpha'=\frac{-\log \sigma^2}{\log P}
\end{equation}
Essentially, the parameter $\alpha'$ measures the quality of the current channel estimation. If $\alpha'=0$, then it corresponds to the case when there is no current CSI. If $\alpha' \geq 1$, then it corresponds to the case that the current CSI is as good as perfect (for DoF). Note that in \cite{Kobayashi_Yang}, the definition of $\alpha'$ is slightly different from our definition where $\alpha'$ is defined as $\lim_{P\rightarrow \infty}\frac{-\log \sigma^2}{\log P}$.

There are two independent messages, one for each receiver. We denote the size of message $W_k$ as $|W_k|$, $k\in \{1,2\}$. For the codewords spanning $n$ channel uses, the rates $R_k=\log(|W_k|)/n$ are achievable if the probability of error for both messages can be simultaneously made arbitrarily small by choosing an appropriately large $n$. The sum capacity $C_{\Sigma}(P)$ is the maximum achievable sum rate. The number of degrees of freedom is defined as
\begin{equation}
d=\lim_{P\rightarrow \infty}\frac{C_{\Sigma}(P)}{\log P}
\end{equation}

Note that there is a subtle thematic distinction between this work and that of Kobayashi et al. in that our focus is limited to optimal use of current and outdated channel information whereas Kobayashi et al. also incorporate temporal correlations. The thematic distinction leads to slightly different channel models. While Kobayashi et al. deal with additional complexities of temporal correlations and non-ergodic settings, we are able to  ignore temporal correlations (i.i.d. fading in time) while capturing the essential aspect of availability of both current and outdated channel knowledge. While our model is simpler than \cite{Kobayashi_Yang} in ignoring temporal correlations, and closer to the original model of \cite{Maddah_Tse}, we expect that from a DoF perspective the two models will produce equivalent results and shed similar insights into the same essential question that motivates the two works.

Next we justify the equivalence of the model presented above (along the lines of Kobayashi et al.) and the simpler model in (\ref{eq:model1}), (\ref{eq:model2}) used for the overview in the introduction section.


\subsection{An equivalent model}\label{sec:simplemodel}
We can perform a sequence of invertible operations at transmitter and receivers to convert the channel to its simplest form. Since invertible transformations do not affect the DoF of the channel, there is no loss of generality, i.e., the resulting channel has the same DoF as the original channel. First, we perform an invertible linear transformation at the transmitter. This is done by multiplying a $2\times 2$ invertible matrix $\mathbf{A}(t)=[\mathbf{v}(t)~ \mathbf{u}(t)]$ to the transmitted signal, i.e., $\mathbf{A}(t)\mathbf{x}(t)$ where $\mathbf{x}(t)=[x_1(t)~x_2(t)]^T$ is the original channel input vector. $\mathbf{v}(t)$ and $\mathbf{u}(t)$ are unit norm vectors chosen in a manner that they are orthogonal to $\hat{\mathbf{g}}(t)$ and $\hat{\mathbf{h}}(t)$, respectively. As a consequence, the received signals at receiver 1 and 2 become
\begin{eqnarray}
y_1(t)&=&\mathbf{h}^\dagger(t)\mathbf{v}(t) x_1(t)+\tilde{\mathbf{h}}^\dagger(t) \mathbf{u}(t) x_2(t)+z_1(t)\\
y_2(t)&=&\mathbf{g}^\dagger(t)\mathbf{u}(t) x_2(t)+\tilde{\mathbf{g}}^\dagger(t) \mathbf{v}(t) x_1(t)+z_2(t)
\end{eqnarray}
Then receiver $i$ can normalize the channel coefficient of $x_i(t)$ to unity such that the channel becomes
\begin{eqnarray}
\underbrace{\frac{y_1(t)}{\mathbf{h}^\dagger(t)\mathbf{v}(t)}}_{y'_1(t)}&=&x_1(t)+\frac{\tilde{\mathbf{h}}^\dagger (t) \mathbf{u}(t)}{\mathbf{h}^\dagger(t)\mathbf{v}(t) } x_2(t)+\underbrace{\frac{z_1(t)}{\mathbf{h}^\dagger(t)\mathbf{v}(t)}}_{z'_1(t)}\\
\underbrace{\frac{y_2(t)}{\mathbf{g}^\dagger(t)\mathbf{u}(t)}}_{y'_2(t)}&=& x_2(t)+\frac{\tilde{\mathbf{g}}^\dagger(t) \mathbf{v}(t)}{\mathbf{g}^\dagger(t)\mathbf{u}(t)} x_1(t)+\underbrace{\frac{z_2(t)}{\mathbf{g}^\dagger(t)\mathbf{u}(t)}}_{z'_2(t)}
\end{eqnarray}
Now let us define $\tilde{\underline{\mathbf{h}}}(t)=\tilde{\mathbf{h}}(t)/\sqrt{P^{-\alpha'}}$ and $\tilde{\underline{\mathbf{g}}}(t)=\tilde{\mathbf{g}}(t)/\sqrt{P^{-\alpha'}}$, both of which have covariance matrix  $\mathbf{I}$. In addition, we normalize the transmit power to unity by absorbing it into the channel coefficients and defining $x_i(t)=\sqrt{P}x'_i(t)$. Then the channel input-output relationship can be written as
\begin{eqnarray}
y'_1(t)&=& \sqrt{P}x'_1(t)+\sqrt{P^{1-\alpha'}}\underbrace{\frac{\tilde{\underline{\mathbf{h}}}^\dagger(t) \mathbf{u}(t)}{\mathbf{h}^\dagger(t)\mathbf{v}(t)}}_{h(t)} x'_2(t)+z'_1(t)\\
y'_2(t)&=& \sqrt{P}x'_2(t)+\sqrt{P^{1-\alpha'}}\underbrace{\frac{\tilde{\underline{\mathbf{g}}}^\dagger (t) \mathbf{u}(t)}{\mathbf{g}^\dagger(t)\mathbf{u}(t)}}_{g(t)} x'_1(t)+z'_2(t)
\end{eqnarray}
where $h(t)$ and $g(t)$ have finite variances independent of $P$  and they are only known to the transmitter with one time delay. By setting $\alpha=1-\alpha'$ and with a little bit of abuse of notations, we end up with the following simple channel model:
\begin{eqnarray}\label{eqn:simplemodel}
y_1(t)&=&\sqrt{P}x_1(t)+\sqrt{P^\alpha}h(t) x_2(t)+z_1(t)\\
y_2(t)&=&\sqrt{P}x_2(t)+\sqrt{P^\alpha}g(t) x_1(t)+z_2(t)
\end{eqnarray}
Since  $\alpha'$ is defined to be greater than or equal to zero and  $\alpha=1-\alpha'$,  it follows that $\alpha \leq 1$.
\section{Results}
When $\alpha<0$, the problem becomes trivial. It can be easily seen that simply zero-forcing at the transmitter can achieve 2 DoF which is also the maximum achievable DoF. Therefore, we only consider the case when $0\leq \alpha \leq 1$ in this section. The main result of this paper is presented in the following theorem.
\begin{theorem}\label{thm:innerbound}
For the MISO broadcast channel with mixed CSIT defined in Section \ref{sec:systemmodel},
\begin{equation}
d = 2-\frac{2}{3}\alpha, \quad 0\leq \alpha \leq 1.
\end{equation}
\end{theorem}
This theorem establishes both the achievable DoF and its optimality. In terms of outer bounds, this is the first non-trivial outer bound for the mixed CSIT setting. However, the result is interesting even in the original setting of \cite{Maddah_Tse} with only delayed CSIT, i.e., with no current CSIT. This is because our outer bound does \emph{not} require that the two users are statistically equivalent, i.e., the users can have different fading distributions. In contrast, the outer bounds for BC with delayed CSIT, e.g., in \cite{Maddah_Tse}, require that two users are statistically equivalent. Therefore, evidently we have a stronger result in the original setting as well.

\section{Outer bounds}
In this section, we provide the outer bound proof. We start first with the original setting of \cite{Maddah_Tse} with only delayed CSIT and provide an alternate proof that does not require statistically equivalent fading for the two users. Our outer bound follows a compound channel approach.
\subsection{Outer bound for the delayed CSIT setting}
Consider the following BC with delayed CSIT as in \cite{Maddah_Tse}.
\begin{eqnarray}
y_1&=&\mathbf{h}^{\dagger}\mathbf{x}+z_1\\
y_2&=&\mathbf{g}^{\dagger}\mathbf{x}+z_2
\end{eqnarray}
For simplicity we omit the time index. Recall that we assume the channels are i.i.d. in time and independent across users but not necessarily identically distributed across users. Now suppose we provide $y_1$ to User 2 so that it has both $y_1, y_2$, which makes the channel physically degraded. For a physically degraded BC without memory (this requirement of memoryless channels is the primary reason that we restrict the model to i.i.d. fading in time), feedback does not increase capacity, so we can eliminate the delayed CSIT feedback for this new channel. Now, let us impose a compound setting on $\mathbf{h}$ and $\mathbf{g}$, which is consistent with the outer bound argument, i.e. the compound setting does not decrease the capacity of the original channel. To see this, suppose we first introduce another pair of receivers, one for each user, that are statistically equivalent to the original receivers and require the same messages. Since the additional receivers have the same decoding capabilities as the original receivers, the capacity region is not decreased. Now we provide full channel knowledge to the transmitter, which  also cannot reduce capacity (this step is not necessary, but it shows the strength of the outer bound). This puts us into a two state compound BC setting. Since at no point did we reduce the capacity, the outer bound for this compound setting is also an outer bound for the original channel.

So now we have two more fictional outputs:
\begin{eqnarray*}
y'_1&=&\mathbf{h}'^{\dagger}\mathbf{x}+z'_1\\
y'_2&=&\mathbf{g}'^{\dagger}\mathbf{x}+z'_2
\end{eqnarray*}
Note that in the compound setting the transmitter knows that the channel vectors can be
either $\mathbf{h},\mathbf{g}$ or $\mathbf{h}',\mathbf{g}'$. Essentially now we have two BCs controlled by the same inputs. In the first BC,  User 1 sees $y_1$ and wants message $W_1$, and User 2 sees $y_1, y_2$ and wants message $W_2$. In the second BC,  the first user sees $y_1'$ and wants message $W_1$ and the second user sees $y_1', y_2'$ and wants message $W_2$. To derive the outer bound, we start with the first BC. From Fano's inequality,
\begin{eqnarray}\label{eqn:or11_dcsit}
n R_1 &\leq& I(W_1; y_1^n) +  o(n) \\
&=&h(y_1^n)-h(y_1^n|W_1)+o(n)\\
&\leq& n\log(P) -h(y_1^n|W_1)+n~o(\log(P))+o(n)
\end{eqnarray}
Proceeding similarly with $y_1'$ instead of $y_1$, we have the bound:
\begin{eqnarray}\label{eqn:or12_dcsit}
 nR_1&\leq& n\log(P) -h(y_1'^n|W_1)+n~o(\log(P))+o(n)
\end{eqnarray}
Adding the two we have
\begin{eqnarray}
2nR_1&\leq& 2n \log(P) -h (y_1^n, y_1'^n | W_1) + n~o(\log(P))+o(n)\label{eqn:or13_dcsit}
\end{eqnarray}
Note that from $y_1^n$ and $y_1'^n$, it is possible to invert the channel and to construct $\mathbf{x}$ within bounded variance noise distortion. Therefore, $h (y_1^n, y_1'^n | W_1) =nR_2+n~o(\log(P))+o(n)$. Using this observation, we proceed as follows.
\begin{eqnarray}
nR_1+nR_2&=&I(W_1,W_2; y_1^n, y_1'^n)+n~o(\log(P))+o(n)\\
nR_2&=&I(W_2; y_1^n, y_1'^n|W_1)+n~o(\log(P))+o(n)\\
&=&h(y_1^n, y_1'^n|W_1) - \underbrace{h(y_1^n, y_1'^n|W_1, W_2)}_{n~o(\log(P))} +n~o(\log(P))+o(n)\nonumber\\
&=& h(y_1^n, y_1'^n|W_1) + n~o(\log(P))+o(n)\label{eqn:r2_dcsit}
\end{eqnarray}
Adding \eqref{eqn:or13_dcsit} and \eqref{eqn:r2_dcsit}, and writing it in DoF terms:
\begin{equation}
2d_1+d_2\leq 2
\end{equation}
By symmetry, we can repeat the whole procedure by creating a degraded channel in the other direction (User 2 is degraded) to obtain the bound
\begin{equation}
2d_2+d_1\leq 2
\end{equation}
Adding the two bounds, we have the final DoF outer bound
\begin{equation}
d_1+d_2\leq \frac{4}{3}.
\end{equation}

Evidently, the outer bound applies even if the channel uncertainty at the transmitter is reduced to a one bit uncertainty representing a choice between only two independent realizations of each user's channel. Note that the bound only requires linear independence between the two realizations. Since the channels are drawn from a continuous distribution, this is true with probability 1.

\subsection{Outer bound for the mixed CSIT setting}
We will derive the outer bound based on the equivalent simple model given by equation \eqref{eqn:simplemodel}. Similar to the outer bound for the delayed CSIT setting, we make the channel physically degraded by providing $y_1$ to User 2. Then we can eliminate the delayed CSIT feedback for this new channel since  feedback does not increase capacity for a physically degraded memoryless BC. Let us impose a compound setting on $h, g$, which is consistent with the outer bound argument. Now we have two more fictional outputs:
\begin{eqnarray*}
y'_1&=&\sqrt{P}x_1+\sqrt{P^{\alpha}}h' x_2+z'_1\\
y'_2&=&\sqrt{P}x_2+\sqrt{P^{\alpha}}g' x_1+z'_2
\end{eqnarray*}
Note that in the compound setting the transmitter knows that the cross-coefficients take values
either $h,g$ or $h',g'$. Essentially now we have two BCs controlled by the same inputs. To derive the outer bound, we start with the first BC. From Fano's inequality,
\begin{eqnarray}
n R_1 &\leq& I(W_1; y_1^n) +o(n)\\
&\leq& n\log(P) -h(y_1^n|W_1)+n~o(\log(P))+o(n)
\end{eqnarray}
Proceeding similarly with $y_1'$ instead of $y_1$, we have the bound:
\begin{eqnarray}
 nR_1&\leq& n\log(P) -h(y_1'^n|W_1)+n~o(\log(P))+o(n)
\end{eqnarray}
Adding the two we have
\begin{eqnarray}
2nR_1&\leq& 2 n\log(P) -h(y_1^n|W_1)-h(y_1'^n|W_1)+n~o(\log(P))+o(n)\\
&\leq& 2n \log(P) -h (y_1^n, y_1'^n | W_1) + n~o(\log(P))+o(n)\label{eqn:r1}
\end{eqnarray}

\noindent To bound the remaining entropy term, next we want to show that
\begin{equation}
nR_2\leq h(y_1^n, y_1'^n|W_1)+n(1-\alpha) \log(P)+n~o(\log(P)) +o(n)
\end{equation}

\noindent Note that from $y_1, y_1'$ we can do a change of basis to obtain
\begin{eqnarray}
y_{1,new}&=&\sqrt{P}x_1 + z_1\\
y_{2,new}&=&\sqrt{P}x_2 + z_2+z
\end{eqnarray}
where $z_i \sim\mathcal{CN}(0,\mathcal{O}(1))$, and most importantly $z\sim\mathcal{CN}(0,P^{1-\alpha})$.

So, from $y_1, y_1'$, and $z$ it is possible to obtain $x_1+z_1, x_2+z_2$, i.e., it is possible to invert the channel within bounded variance noise distortion. Using this observation, we proceed as follows.
\begin{eqnarray}
nR_1+nR_2&=&I(W_1,W_2; y_1^n, y_1'^n, z^n)+n~o(\log(P))+o(n)\\
nR_2&=&I(W_2; y_1^n, y_1'^n, z^n|W_1)+n~o(\log(P))+o(n)\\
&=& I(W_2; y_1^n, y_1'^n|W_1)+I(W_2; z^n|y_1^n, y_1'^n, W_1)+n~ o(\log(P))+o(n)\\
&=&h(y_1^n, y_1'^n|W_1) - \underbrace{h(y_1^n, y_1'^n|W_1, W_2)}_{n~o(\log(P))}+ \underbrace{h(z^n|y_1^n, y_1'^n, W_1)}_{\leq h(z^n)}\nonumber\\
&& ~~~-h(z^n|y_1^n, y_1'^n, W_1,W_2) +n~ o(\log(P))+o(n)\nonumber\\
&\leq&h(y_1^n, y_1'^n|W_1) +h(z^n)-\underbrace{h(z^n|(z_2+z)^n)}_{n~o(\log(P))}+n~o(\log(P))+o(n)\\
&=& h(y_1^n, y_1'^n|W_1) + n(1-\alpha)\log(P)+n~o(\log(P))+o(n)\label{eqn:r2}
\end{eqnarray}

\noindent Thus we have found a bound for the remaining entropy term. Adding \eqref{eqn:r1} and \eqref{eqn:r2}, we have
\begin{eqnarray}
2nR_1+nR_2&\leq& 2n\log(P)+n(1-\alpha)\log(P)+n ~o(\log(P))+o(n)
\end{eqnarray}
Writing the equation in DoF terms, we have
\begin{eqnarray}
2d_1+d_2&\leq& 3-\alpha
\end{eqnarray}

\noindent By symmetry we can repeat the whole procedure by creating a degraded channel in the other direction (User 2 is degraded) to obtain the bound:
\begin{eqnarray}
d_1+2d_2&\leq& 3-\alpha
\end{eqnarray}
Adding the two bounds, we have the final DoF outer bound
\begin{eqnarray}
d_1+d_2&\leq& 2-\frac{2}{3}\alpha
\end{eqnarray}
which represents a straight line that goes from a sum DoF value of $d_1+d_2=2$ when $\alpha=0$ to a sum DoF value of $d_1+d_2=4/3$ when $\alpha=1$. 

\section{Achievability}
In this section we provide the achievability proof for Theorem \ref{thm:innerbound}. The achievable scheme consists of three phases. Since the achievable schemes share many similarities with that derived in \cite{Kobayashi_Yang}, we will use the equivalent model given by \eqref{eqn:model} for ease of exposition.

\subsection*{Phase 1}
Phase 1 consists of one time slot and the transmitted signal is
\begin{equation}
\mathbf{x}(1)=\mathbf{v}_1a_1+\mathbf{v}_2a_2+\mathbf{u}_1b_1+\mathbf{u}_2 b_2
\end{equation}
where $a_i$ and $b_i$ are symbols intended for user 1 and 2, respectively. Two independent Gaussian codebooks are used for $[a_1 ~a_2]$ and $[b_1~ b_2]$ with diagonal covariance matrices. The power of $a_1$ and $b_1$ is $(P-P^{1-\alpha'-\epsilon})/2$ where $\epsilon$ is a positive number that can be chosen arbitrarily small while the power of $a_2$ and $b_2$ is $P^{1-\alpha'-\epsilon}/2$ such that $E[\|\mathbf{x}(1)\|^2]\leq P$. $\mathbf{v}_i$ and $\mathbf{u}_i$ are beamforming vectors with unit norm. We will design $\mathbf{v}_1$ and $\mathbf{u}_1$ such that they are orthogonal to $\hat{\mathbf{g}}(1)$ and $\hat{\mathbf{h}}(1)$, respectively, i.e.
\begin{eqnarray}
\hat{\mathbf{g}}^{\dagger}(1)\mathbf{v}_1=0\\
\hat{\mathbf{h}}^{\dagger}(1)\mathbf{u}_1=0
\end{eqnarray}
$\mathbf{v}_2$ and $\mathbf{u}_2$ are chosen randomly, such that they are linearly independent with $\mathbf{v}_1$ and $\mathbf{u}_1$, respectively, with probability one.

\begin{figure*}[!t]
\centerline{\subfigure[Received signal at Receiver 1
]{\includegraphics[width=2.6in]{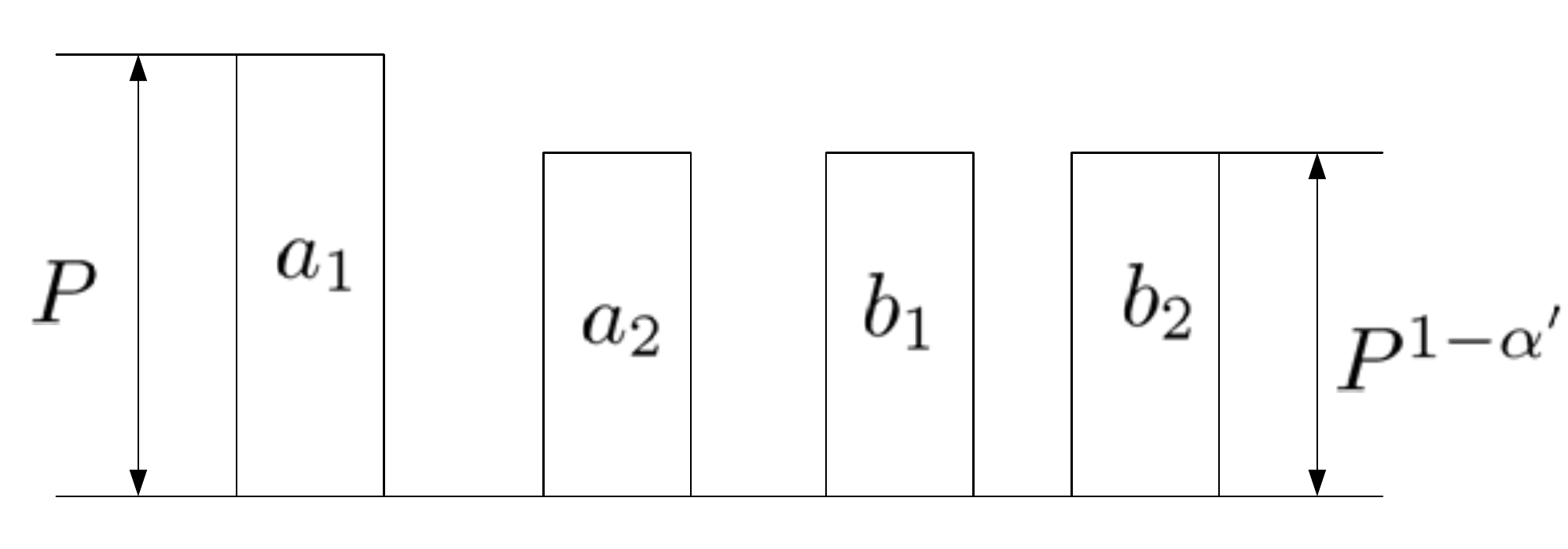} \label{fig:rx1t1}} \hfil \subfigure[Received signal at Receiver 2]{\includegraphics[width=2.6in]{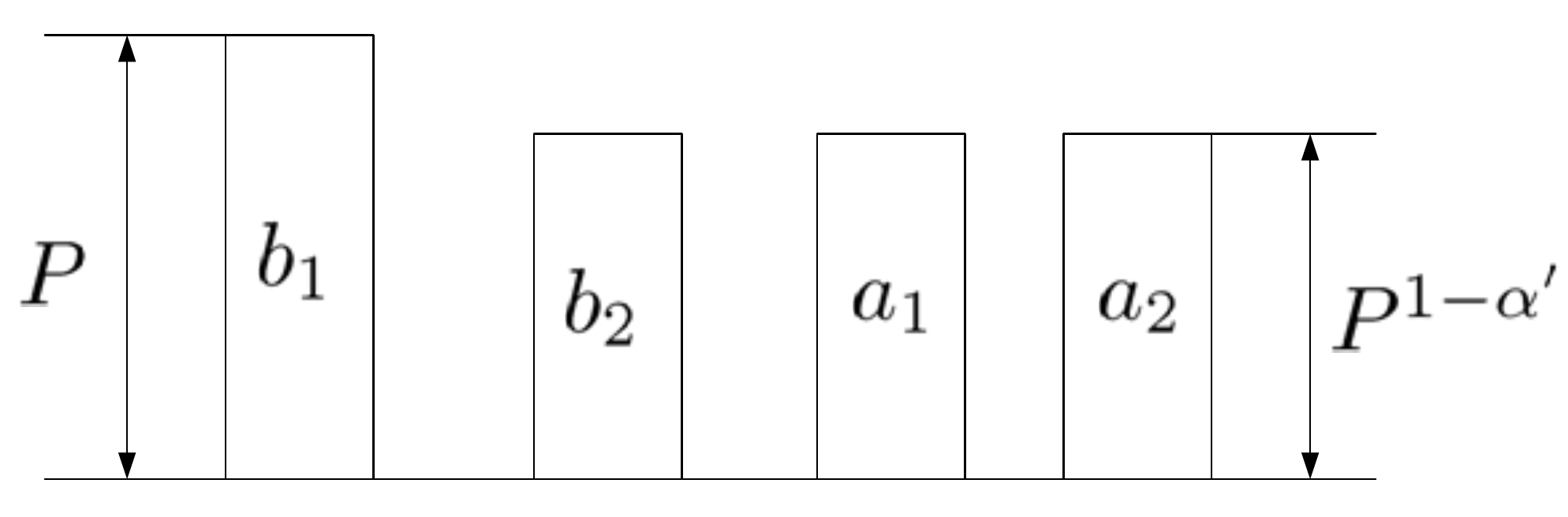}\label{fig:rx2t1} }}
\caption{The power levels of symbols at two receivers in time slot 1}\label{fig:t1}
\end{figure*}

Now consider the received signal at receiver 1.
\begin{eqnarray}
y_1(1)&=&\mathbf{h}^{\dagger}(1)\mathbf{x}(1)+z_1(1)\\
   &=&\mathbf{h}^{\dagger}(1)\mathbf{v}_1a_1+\mathbf{h}^{\dagger}(1)\mathbf{v}_2a_2+\underbrace{\left(\tilde{\mathbf{h}}^{\dagger}(1)\mathbf{u}_1b_1+\mathbf{h}^{\dagger}(1)\mathbf{u}_2b_2\right)}_{\eta_1}+z_1(1)
\end{eqnarray}
Similarly, the received signal at receiver 2 is
\begin{eqnarray}
y_2(1)&=&\mathbf{g}^{\dagger}(1)\mathbf{x}(1)+z_2(1)\\
   &=&\mathbf{g}^{\dagger}(1)\mathbf{u}_1b_1+\mathbf{g}^{\dagger}(1)\mathbf{u}_2b_2+\underbrace{\left(\tilde{\mathbf{g}}^{\dagger}(1)\mathbf{v}_1a_1+\mathbf{g}^{\dagger}(1)\mathbf{v}_2a_2\right)}_{\eta_2}+z_2(1)
\end{eqnarray}
The power level of each symbol at each receiver is shown in Fig.\ref{fig:t1}. As we can see, due to partial zero-forcing of symbol $a_1$ and $b_1$ at receiver 2 and receiver 1, respectively, they are received at power level of $P^{1-\alpha'}$ although they are sent with power level of $O(P)$. On the other hand, $a_2$ and $b_2$ cannot be partially zero-forced at the unintended receivers and we allocate the power such that they are received at the same power level of $a_1$ and $b_1$ at receiver 2 and 1, respectively.

As in \cite{Kobayashi_Yang}, we will quantize the real and imaginary parts of $\eta_k$ separately using a scalar  truncated uniform quantizer with unit step and truncation value $\bar{\eta}=P^{\frac{1+\zeta}{2}}\sigma$, for some $\zeta>0$. Let us denote
\begin{equation}
\eta_k=\hat{\eta}_k+\Delta_k
\end{equation}
where $\hat{\eta}_k$ is the quantized value while $\Delta_k$ is the quantization error. And $\hat{\eta}_k$ contains
\begin{equation}
R_k=2 \log (2\lceil\bar{\eta}\rceil) \approx 2+(1+\zeta-\alpha') \log P\quad \textrm{bits}
\end{equation}
 $\hat{\eta}_k$ viewed as a message containing $R_k$ bits will be encoded using a Gaussian codebook with codewords denoted by $c_k$. Then $c_k$ will be sent to both receivers as a common information in the following time slots. Note that essentially phase 1 is the same as phase 1 of the scheme proposed in \cite{Kobayashi_Yang}.

\subsection*{Phase 2}
\begin{figure*}
\centerline{\subfigure[Received signal at Receiver 1
]{\includegraphics[width=2.4in]{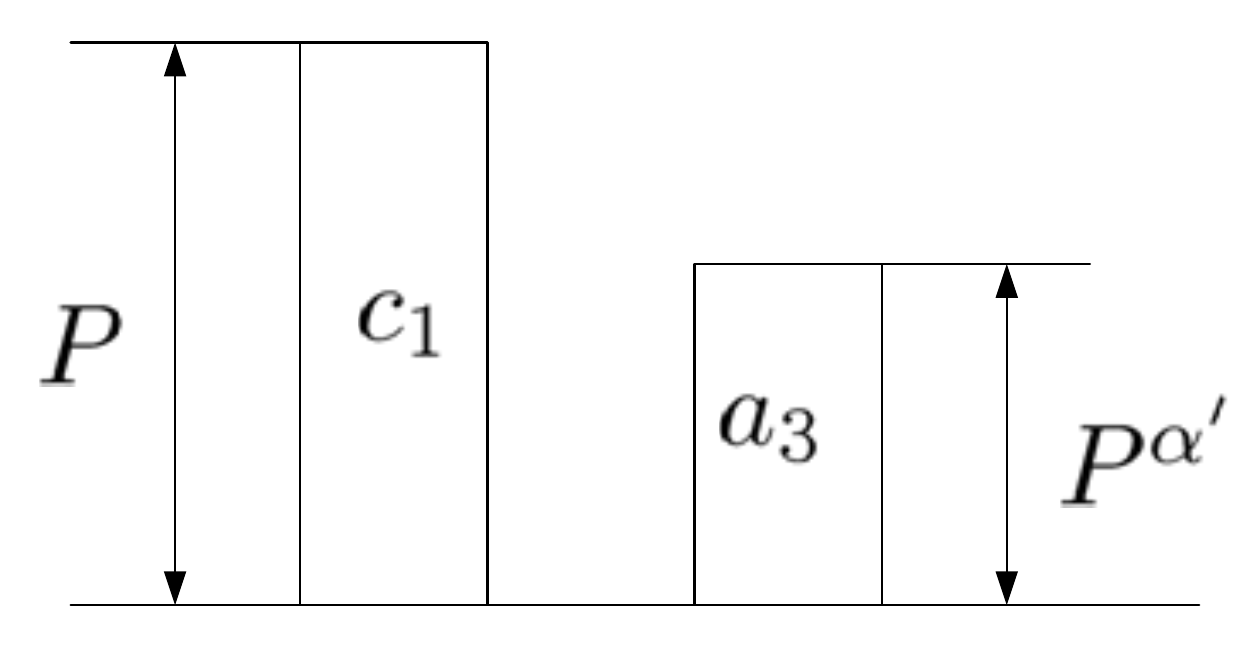} \label{fig:rx1t2}} \hfil \subfigure[Received signal at Receiver 2]{\includegraphics[width=2.4in]{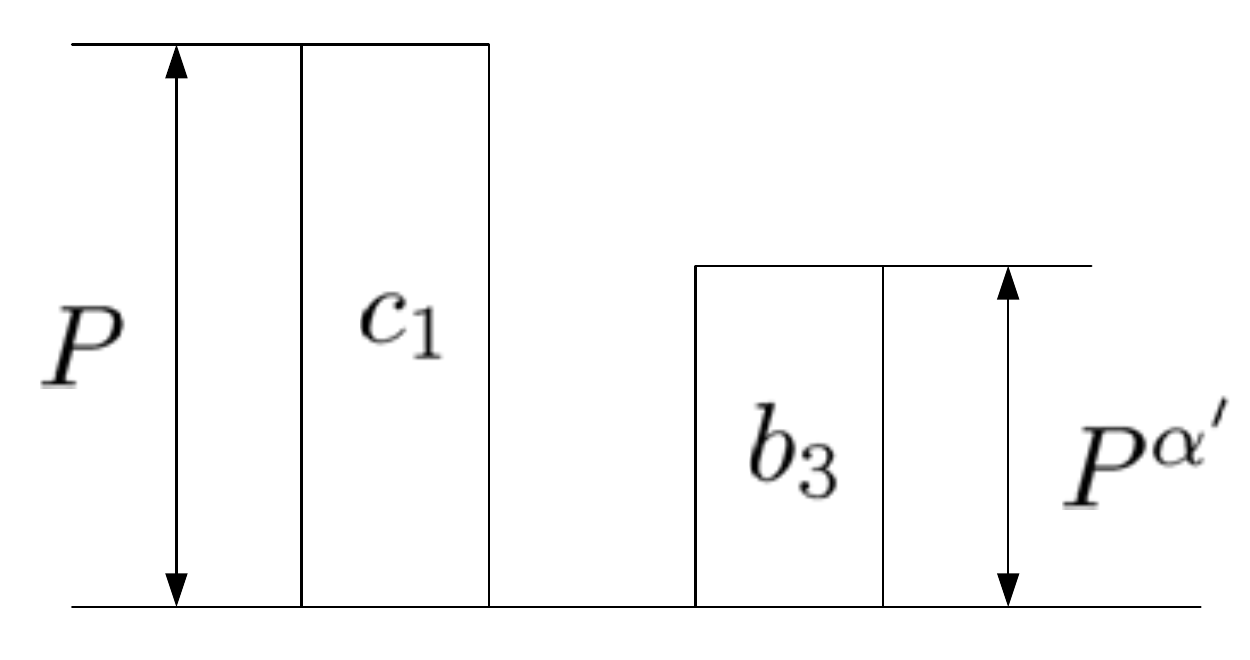}\label{fig:rx2t2} }}
\caption{The power levels of symbols at two receivers in phase 2}\label{fig:t2}
\end{figure*}

The goal of phase 2 is to deliver the common information $c_1$ to both receivers and at the same time to send two private messages each for one receiver. Suppose private messages for receiver 1 and 2 are encoded using independent Gaussian codes with codewords denoted as $a_3$ and $b_3$, respectively. Consider the following transmitted signal (for simplicity we omit the time index)
\begin{equation}
\mathbf{x}=\mathbf{w}c_1+\mathbf{v}_3a_3+\mathbf{u}_3b_3
\end{equation}
where $\mathbf{w}$, $\mathbf{v}_3$ and $\mathbf{u}_3$ are beamforming vectors with unit norm. $\mathbf{w}$ is chosen randomly while $\mathbf{v}_3$ and $\mathbf{u}_3$ are chosen such that they are orthogonal to $\hat{\mathbf{g}}$ and $\hat{\mathbf{h}}$, respectively. We set the powers of $a_3$, $b_3$ and $c_1$ as $P^{\alpha'}/2$, $P^{\alpha'}/2$ and $P-P^{\alpha'}$, respectively, such that $E[\|\mathbf{x}\|^2]=P$. Then the received signal at receiver 1 is
\begin{eqnarray}
y_1&=&\mathbf{h}^{\dagger}\mathbf{x}+z_1\\
   &=&\mathbf{h}^{\dagger}\mathbf{w}c_1+\mathbf{h}^{\dagger}\mathbf{v}_3a_3+\tilde{\mathbf{h}}^{\dagger}\mathbf{u}_3b_3+z_1.
\end{eqnarray}
Now consider the power level of each symbol at receiver 1. It can be easily seen that $E[|\mathbf{h}^{\dagger}\mathbf{w}c_1|^2]$ and $E[|\mathbf{h}^{\dagger}\mathbf{v}_3a_3|^2]$ are on the order of $P$ and $P^{\alpha'}$, respectively, as shown in Fig. \ref{fig:rx1t2}. For $b_3$, it can be seen that $E[|\tilde{\mathbf{h}}^{\dagger}\mathbf{u}_3b_3|^2]$ is $O(1)$, i.e. at the noise floor level.
Similarly, the received signal at receiver 2 is
\begin{eqnarray}
y_2&=&\mathbf{g}^{\dagger}\mathbf{x}+z_2\\
   &=&\mathbf{g}^{\dagger}\mathbf{w}c_1+\mathbf{g}^{\dagger}\mathbf{u}_3b_3+\tilde{\mathbf{g}}^{\dagger}\mathbf{v}_3a_3+z_2
\end{eqnarray}
Again, $a_3$ is received at the noise floor level while the power levels of $c_1$ and $b_3$ are $P$ and $P^{\alpha'}$, respectively, as shown in Fig. \ref{fig:rx2t2}.
Both receivers will first decode $c_1$ by treating other signals as noise. The achievable rate for $c_1$ is
\begin{eqnarray}
R_{c_1}&=&\min\left\{I(c_1;y_1,\mathbf{h},\mathbf{g}),I(c_1;y_2,\mathbf{h},\mathbf{g})\right\}
\end{eqnarray}
Next, we calculate these two mutual information terms. First,
\begin{eqnarray}
I(c_1;y_1,\mathbf{h},\mathbf{g})&=&E\left[\log\left(1+\frac{|\mathbf{h}^{\dagger}\mathbf{w}|^2(P-P^{\alpha'})}{1+|\mathbf{h}^{\dagger}\mathbf{v}_3|^2\frac{P^{\alpha'}}{2}+|\tilde{\mathbf{g}}^{\dagger}\mathbf{u}_3|^2\frac{P^{\alpha'}}{2}}\right)\right]\\
&=&E\left[\log\left(1+\frac{|\mathbf{h}^{\dagger}\mathbf{w}|^2(P-P^{\alpha'})}{1+|\mathbf{h}^{\dagger}\mathbf{v}_3|^2\frac{P^{\alpha'}}{2}+|P^{\frac{-\alpha'}{2}}\tilde{\underline{\mathbf{g}}}^{\dagger}\mathbf{u}_3|^2\frac{P^{\alpha'}}{2}}\right)\right]\\
&=&E\left[\log\left(1+\frac{|\mathbf{h}^{\dagger}\mathbf{w}|^2(P-P^{\alpha'})}{1+|\mathbf{h}^{\dagger}\mathbf{v}_3|^2\frac{P^{\alpha'}}{2}+|\tilde{\underline{\mathbf{g}}}^{\dagger}\mathbf{u}_3|^2/2}\right)\right]\\
&=& E\left[\log\left(\frac{|\mathbf{h}^{\dagger}\mathbf{w}|^2(P-P^{\alpha'})}{|\mathbf{h}^{\dagger}\mathbf{v}_3|^2\frac{P^{\alpha'}}{2}}\right)\right]+o(\log P)\\
&=&E\left[\log\left(\frac{2|\mathbf{h}^{\dagger}\mathbf{w}|^2(P^{1-\alpha'}-1)}{|\mathbf{h}^{\dagger}\mathbf{v}_3|^2}\right)\right]+o(\log P)\\
&=& (1-\alpha')\log P + o(\log P)
\end{eqnarray}
where $\tilde{\underline{\mathbf{g}}}=\tilde{\mathbf{g}}/P^{\frac{-\alpha'}{2}}$ with covariance matrix $\mathbf{I}$. By symmetry, we also have
\begin{eqnarray}
I(c_1;y_2,\mathbf{h},\mathbf{g})=(1-\alpha')\log P + o(\log P)
\end{eqnarray}
Therefore, $R_{c_1}=(1-\alpha')\log P +o(\log P)$. Note that $c_1$ contains $2+(1+\zeta-\alpha') \log P$ bits.  The number of time slots needed to send these bits is
\begin{equation}
t_2=\frac{2+(1+\zeta-\alpha') \log P}{(1-\alpha')\log P+o(\log P)}=\frac{2}{(1-\alpha')\log P+o(\log P)}+\frac{(1+\zeta-\alpha')}{1-\alpha'+\frac{o(\log P)}{\log P}}.
\end{equation}
Since $\zeta$ can be made arbitrarily small,  $t_2\rightarrow 1$  as $P\rightarrow \infty$. In other words, at high SNR, phase 2 only consists of one time slot.

After decoding $c_1$ at receiver 1, since the channels are known to the receiver, it can remove $c_1$ from the received signal. Then it can decode $a_3$ which  achieves the following rate:
\begin{eqnarray}
R_{a_3}&=& I(a_3;y_2,\mathbf{h},\mathbf{g}|c_1)\\
&=&E\left[\log\left(1+\frac{|\mathbf{h}^{\dagger}\mathbf{v}_1|^2\frac{P^{\alpha'}}{2}}{1+|\tilde{\mathbf{g}}^{\dagger}\mathbf{v}_1|^2\frac{P^{\alpha'}}{2}}\right)\right]\\
&=&E\left[\log\left(1+\frac{|\mathbf{h}^{\dagger}\mathbf{v}_1|^2\frac{P^{\alpha'}}{2}}{1+|\tilde{\mathbf{\underline{g}}}^{\dagger}\mathbf{v}_1|^2/2}\right)\right]\\
&=& \alpha' \log P+ o(\log P)
\end{eqnarray}
By symmetry, at receiver 2, $b_3$ can achieve a rate
\begin{equation}
R_{b_3}=I(b_3;y_2,\mathbf{h},\mathbf{g}|c_1)= \alpha' \log P+o(\log P)
\end{equation}

\subsection*{Phase 3}
Phase 3 is very similar to Phase 2. In phase 3, the common information $c_2$ will be sent to both receivers and again at the same time one private message will be sent to each receiver. Denote the codewords of private message for receiver 1 and 2 as $a_4$ and $b_4$, respectively. Then the transmitted signal is
\begin{equation}
\mathbf{x}=\mathbf{w}c_2+\mathbf{v}_4a_4+\mathbf{u}_4b_4
\end{equation}
where  $\mathbf{v}_4$ and $\mathbf{u}_4$ with unit norm  are chosen such that they are orthogonal to $\hat{\mathbf{g}}$ and $\hat{\mathbf{h}}$, respectively. We set the powers of $a_4$, $b_4$ and $c_2$ as $P^{\alpha'}/2$, $P^{\alpha'}/2$ and $P-P^{\alpha'}$, respectively. With the same analysis as phase 2,  both $a_4$ and $b_4$ can achieve $\alpha'$  DoF  and $c_2$ can be delivered using $t_3$ time slots where $t_3=t_2$.

After decoding $c_1$ and $c_2$, both receivers know  $\hat{\eta}_1$ and $\hat{\eta}_2$. Now receiver 1 will use $\hat{\eta}_1$, $\hat{\eta}_2$ and $y_1(1)$ to construct an effective $2\times2$ MIMO channel with input $a_1$ and $a_2$ as follows.
\begin{eqnarray}
\left[\begin{array}{c}y_1(1)-\hat{\eta}_1\\ \hat{\eta}_2 \end{array}\right]=\left[\begin{array}{c}\mathbf{h}^{\dagger}(1)\\ \mathbf{g}^{\dagger}(1)\end{array}\right]\left[\begin{array}{cc}\mathbf{v}_1& \mathbf{v}_2\end{array}\right]\left[\begin{array}{c}a_1\\ a_2 \end{array}\right]+\left[\begin{array}{c}z_1(1)+\Delta_1\\ -\Delta_2 \end{array}\right]
\end{eqnarray}
Similarly, receiver 2 will use $\hat{\eta}_1$, $\hat{\eta}_2$ and $y_2(1)$ to construct an effective MIMO channel with input $b_1$ and $b_2$ as follows.
\begin{eqnarray}
\left[\begin{array}{c}y_2(1)-\hat{\eta}_2\\ \hat{\eta}_1 \end{array}\right]=\left[\begin{array}{c}\mathbf{g}^{\dagger}(1)\\ \mathbf{h}^{\dagger}(1)\end{array}\right]\left[\begin{array}{cc}\mathbf{u}_1& \mathbf{u}_2\end{array}\right]\left[\begin{array}{c}b_1\\ b_2 \end{array}\right]+\left[\begin{array}{c}z_2(1)+\Delta_2\\ -\Delta_1 \end{array}\right]
\end{eqnarray}
As shown in \cite{Kobayashi_Yang}, the probability of error $P_e$ of decoding $a_1$, $a_2$, $b_1$ and $b_2$ will go to zero as $P\rightarrow \infty$. And the rates of $[a_1~ a_2]$ and $[b_1 ~b_2]$  are $(2-\alpha')\log P +o(\log P)$. Now, using random coding arguments, it can be seen for $[a_1~ a_2]$ and $[b_1 ~b_2]$, the reliable rates are approximately $(1-P_e)((2-\alpha')\log P +o(\log P))$. Since $P_e$ will go to zero when $P \rightarrow \infty$, $[a_1~ a_2]$ and $[b_1~ b_2]$ can achieve $2-\alpha'$ DoF.

Now we can calculate the achievable DoF using this scheme. When $P\rightarrow \infty$, phase 2 and 3 each consists of 1 time slot. Since phase 1 consists of 1 time slot, a total of 3 time slots are used. Over these 3 time slots, four symbols are delivered to each receiver. Therefore,
\begin{equation}
d= \frac{2(2-\alpha'+\alpha'+\alpha')}{3}=\frac{2(2+\alpha')}{3}.
\end{equation}
Replacing $\alpha'$ with $1-\alpha$, we have
\begin{equation}
d=2-\frac{2}{3}\alpha.
\end{equation}

\section{Conclusion}
We consider a two user MISO BC with two antennas at the transmitter. It is assumed the transmitter has delayed CSI and imperfect current CSI. We characterize the optimal DoF of this channel by providing both achievability and outer bounds. The results reveal the DoF optimal use of  outdated  and current CSIT.

\end{document}